\documentclass[twocolumn, aps,pre,groupedaddress,showpacs]{revtex4-1}
\usepackage{amsfonts}
\usepackage{amsmath}
\usepackage{amssymb}
\usepackage{txfonts}
\usepackage{pxfonts}
\usepackage{graphicx,bm,units,yfonts}
\usepackage[table]{xcolor}
\usepackage{hyperref}

\newcommand{\CM}{{\mathbb C}}

\newcommand{\TM}{{\mathbb T}}
\newcommand{\ZM}{{\mathbb Z}}

\begin{document}

\title{Valley-Chern Effect with LC-Resonators: A Modular Platform}

\author{Yishai Eisenberg}
\affiliation{Department of Physics, Yeshiva University, New York, NY, USA}

\author{Yafis Barlas}
\affiliation{Department of Physics, Yeshiva University, New York, NY, USA}

\author{Emil Prodan}
\affiliation{Department of Physics, Yeshiva University, New York, NY, USA}

\email{prodan@yu.edu}

\begin{abstract}
The valley Chern-effect is theoretically demonstrated with a novel alternating current circuitry, where closed-loop LC-resonators sitting at the nodes of a honeycomb lattice are inductively coupled along the bonds. This enables us to generate a dynamical matrix which copies identically the Hamiltonian driving the electrons in graphene. The valley-Chern effect is generated by splitting the inversion symmetry of the lattice. After a detailed study of the Berry curvature landscape and of the localization of the interface modes, we derive an optimal configuration of the circuit. Furthermore, we show that Q-factors as high as $10^4$ can be achieved with reasonable materials and configurations.
\end{abstract}

\pacs{03.65.Vf, 05.30.Rt, 71.55.Jv, 73.21.Hb}

\maketitle


\section{Introduction}

The physics of a honeycomb lattice is often determined by two small pockets of the Brillouin zone, which are referred to as the valleys. This valley degree of freedom can be controlled and manipulated like the spin~\cite{XiaoPRL2007,YaoPRB2008}. Due to time-reversal and inversion symmetry of a honeycomb lattice, the dispersion at these valleys is gapless and linear, similar to that of massless Dirac fermions. When the energy spectrum becomes gapped as a result of breaking the inversion-symmetry of the honeycomb lattice, a unique topological effect emerges~\cite{FirozIslamCarbon2016,RenRPP2016}. This effect results from a Berry curvature accumulation at the valleys  and is manifested as the emergence of counter-propagating quasi-chiral modes along an interface separating two mirror-inverted asymmetric honeycomb lattices. Since the effect does not rely on a fermionic time-reversal symmetry or on breaking of such symmetry, there is no barrier to realization of this effect in generic dynamical systems. In fact, this valley-Chern effect has been already been demonstrated in photonic~\cite{MaNJP2016,ChenArxiv2016,DongNatMat2017,ChenPRB2017,
BleuPRB2017,NiSciAdv2018,GaoPRB2018,GaoNatPhys2018,
NohPRL2018,KangNatComm2018} and phononic (condensed atomic matter or continuum mechanics) ~\cite{ZhuArxiv2017,JiangArxiv2017,LiuPRA2018,WuSciRep2018,
ChaunsaliPRB2018,ChernArxiv2018} systems. Similarly, meta-materials emulating  mechanical versions of graphene have been engineered to realize the valley-Chern effect with mechanical resonators~\cite{PalNJP2017,VilaPRB2017,QianPRB2018} and acoustic cavities~\cite{LuPRL2016,LuNatPhys2017}. 

Practical applications of the valley-Chern effect are still in the future. The main obstacle in harnessing the topological wave-guiding supplied by the effect is the modest Q-factors of the platforms used so far. Therefore, the search for high Q-factor implementations continues. Such implementations will not only enable practical applications but will also give us access to the unique fundamental physics of topological phenomena, such as the critical behavior at a topological Anderson localization-delocalization transition. Regarding the latter, let us point out that the Q-factor determines an effective size where coherent phenomena can be observed, very much like Thouless' effective length \cite{ThoulessPRL1977} for dissipative electronic systems. Since Anderson's localization length diverges at a topological transition, resolving the critical regime requires large effective sizes, hence large Q-factors. For similar reasons, proving delocalization of edge or interface modes require extremely large Q-factors. 

\vspace{0.2cm}
In this paper, we propose an electrical realization of the valley-Chern effect on a honeycomb lattice, using an inductively coupled network of LC resonators. Topological circuits~\cite{tcircuit1,tcircuit2} emulating topologically non-trivial hopping matrices have been designed with one-~\cite{LCZakphase, tcircuit3}, two-~\cite{tcircuit1,tcircuit2,LeiblatticeLC} three-~\cite{WeylLC} dimensional lattices. In particular, LC electric circuits networks can be designed to yield topological phases and topological phase transitions by simply controlling variable capacitors. However, up to now, all topological circuits either consist of a single, or collection of capacitors, at one site connected to another site either by inductors or resistors. The latter simulate hopping terms while the capacitors provide the local degrees of freedom. These designs, however, cannot be pictured as coupled resonators because the circuits do not support self-sustained signals once the bonds are eliminated. This makes it difficult to generate circuits which emulate generic tight-binding Hamiltonians. In contradistinction, our design consists of closed-loop LC-resonators coupled to each other inductively, by threading adjacent inductors with a ferromagnetic ring as in a transformer. Contrary to earlier proposals, the current in each LC resonator is conserved and corresponds to a local site variable. The difference in this LC circuit network is that each LC resonator can self-sustain a current when the bonds are cut and, furthermore, the resonators maintain their identities even when the bonds present. As such, the circuit can be pictured as a network of well defined coupled resonators, hence it can be directly mapped onto a tight-binding hopping matrix Hamiltonian \cite{ourpaper}. The discrete nature of each LC resonator and it's variable inductive coupling provides a versatile modular platform, that can assist in realization of other topological phases of matter~\cite{ourpaper} by a direct patterning of the mutual inductances. 

\begin{figure}
\center
  \includegraphics[width=0.9\linewidth]{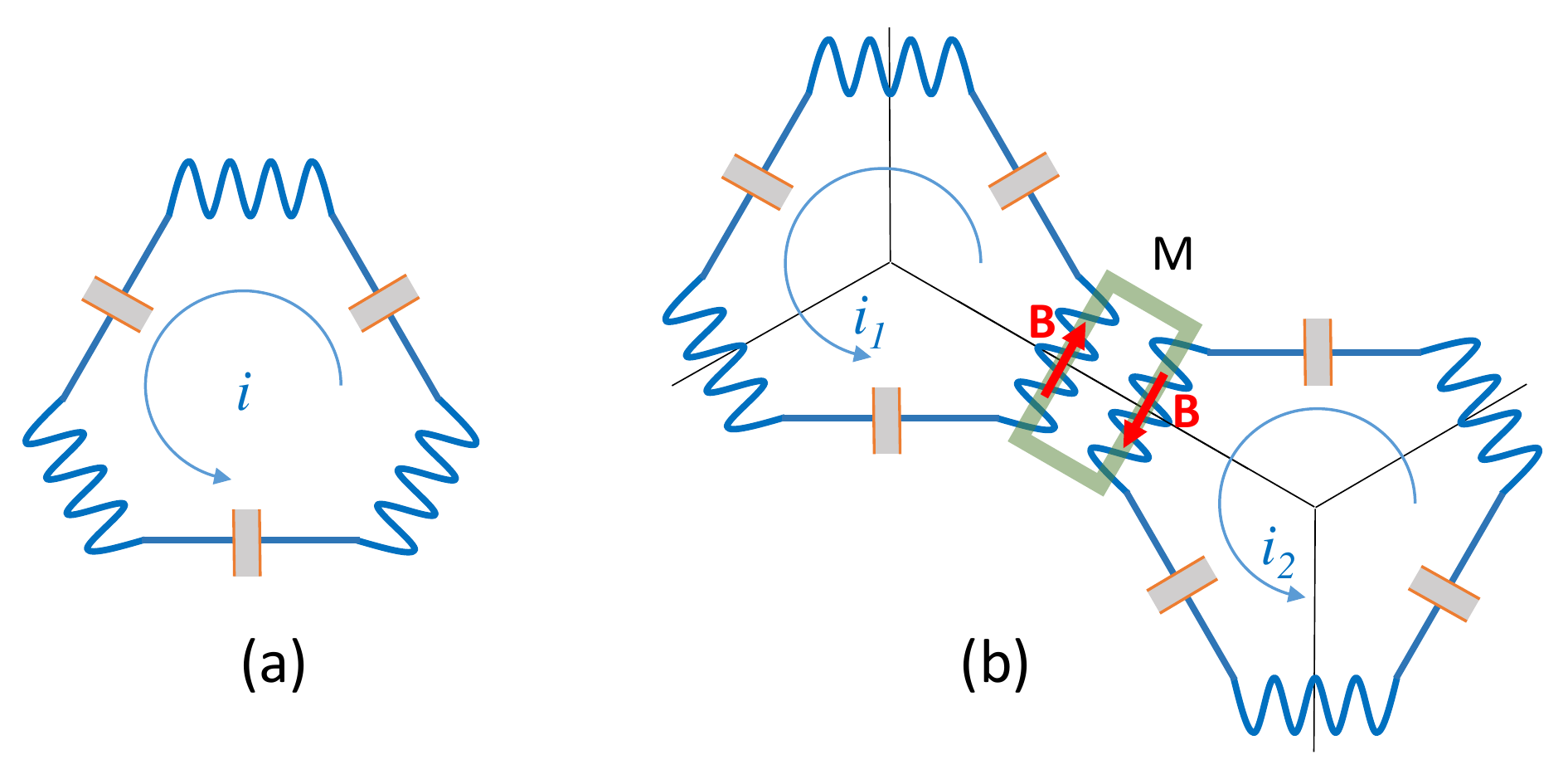}\\
  \caption{\small Basic resonators and their coupling. (a) The resonators consist of a closed loop-circuit containing 3 identical capacitors and 3 identical inductors connected in series. This gives the circuit a 3-fold symmetry, which is used to arrange the resonators in a honeycomb lattice. The resonators have a single degree of freedom, which is the electric current, whose positive orientation is considered counter-clockwise. (b) Two resonators are coupled inductively through a ferromagnetic ring threaded trough adjacent inductors.}
 \label{Fig:Coupling}
\end{figure}

To realize the valley Chern effect, we place each LC resonator at the vertex of a honeycomb lattice and connect it inductively to its nearest neighbors, hence simulating the links of a honeycomb lattice. Each LC resonator consists three capacitors and inductors arranged in a triangular geometry (see Fig~\ref{Fig:Coupling}). The valley-Chern effect is induced by assigning different values to the total capacitance for any near neighboring LC resonator in the honeycomb lattice, while keeping all the mutual and self-inductances same through out the lattice, thereby breaking the inversion symmetry. The inversion asymmetry can be captured by a single parameter $r$, which is proportional to the difference in the total capacitance at near neighboring sites. When $r=0$, the frequency spectrum consists of the two dispersive bands touching at two points, where conical singularities occur, similar to the Dirac cones in a honeycomb lattice. However, when $r > 0$ the Dirac cones split and a spectral gap in the frequency spectrum emerges, resulting in a non-zero Berry curvature. We show that when a domain wall is designed, by choosing the total capacitance of near neighbors along the domain wall the be equal, one-dimensional counter-propagating wave guiding modes emerge along the interface. Using realistic parameters, we show that these LC circuits can be tuned to have large Q-factors in the range $\sim 10^3-10^4$. Such large Q-factors in topological meta-materials can assist in detection and observation of fundamental physics, such as the topological Anderson localization-delocalization transition. 

The rest of the paper is organized as follows: In section II and III, we describe the basic LC resonators which can be inductively coupled  simulating the inter-site coupling (or hopping) on the two-dimensional planar lattice and solve the circuit equations for the honeycomb lattice of LC resonators. In section IV, we calculate the frequency dispersion and topological properties of the electrical lattice as a function of $r$, quantifying the degree of inversion asymmetry of the topological circuit. In section V, we describe how to engineer domain wall configurations which can support counter-propagating one-dimensional wave guiding modes along the domain wall. In the last section, we comment on the experimental realization of the circuit and analyze parameters which result in large $Q$-factors.

\section{The basic resonators and their couplings}

The basic building blocks for the proposed extended planar circuits are single-mode electrical resonators as the one shown in Fig.~\ref{Fig:Coupling}(a). The electrical resonator is a closed LC-circuit with 3-fold symmetry, which will enable us to assemble them into a honeycomb lattice. The electrical time-variable current flowing through this circuit will represent the degree of freedom of the resonator. Throughout, the positive flow will always be counter-clockwise and the symbols like $L$, $C$, $q$ etc. will represent the net inductance, capacitance, charge, etc. in a closed loop circuit.

The novelty of our proposal is the inductive coupling of the resonators shown in Fig.~\ref{Fig:Coupling}(b). In this diagram, the inductive coupling is realized through a ferromagnetic ring threading two neighboring inductors, very much like one will find in a transformer. Since the currents are conserved for each loop of the circuit, the degrees of freedom of our resonators remain well defined and the circuits can be treated as coupled discrete resonators. 

For the coupling in Fig.~\ref{Fig:Coupling}(b), Kirchhoff's  law leads to the system of coupled equations:
\begin{align}\label{Eq:Coupling1}
& \frac{Q_1}{C_1} + L_1 \frac{dI_1}{dt} + M \frac{d I_2}{dt} = 0, \\
& \frac{Q_2}{C_2} + L_2 \frac{dI_2}{dt} + M \frac{d I_1}{dt} = 0.
\end{align}
Using the Fourier decomposition:
\begin{equation}
I_\alpha(t) = \int_{-\infty}^{\infty} {\rm d} \omega \, i_\alpha(\omega) e^{j\omega t} , \quad \alpha=1,2,
\end{equation}
together with $i_\alpha(-\omega) = i_\alpha(\omega)^\ast$, which ensures that $I_\alpha$ are real quantities, we write:
\begin{equation}\label{Eq-Fourier}
I_\alpha(t) = {\rm Re}\Big [\int_{0}^{\infty} {\rm d} \omega \,  i_\alpha(\omega)e^{j\omega t} \Big ], \quad \alpha=1,2,
\end{equation}
with no constraints attached to $i_\alpha(\omega)$ as long as $\omega$ is restricted to the positive real line, which we will enforce through out.
Applying the operator $\frac{1}{C_\alpha} \int dt + L_1 \frac{d}{dt}+ M \frac{d}{dt}$ on \eqref{Eq-Fourier}, we obtain the equations for the Fourier amplitudes:
\begin{align}\label{Eq:Coupling2}
& \bigg (\frac{1}{j \omega C_1} + j\omega L_1 \bigg ) \, i_1(\omega) + j \omega M \, i_2(\omega) = 0, \\ \label{Eq:Coupling3}
& \bigg (\frac{1}{j \omega C_2} + j\omega L_2 \bigg ) \, i_2(\omega) + j \omega M \, i_1(\omega) = 0.
\end{align}
Additional coupling terms will be included when the resonators couple to more than one adjacent loop.

The platform we propose here can be generalized and a number of arbitrary inductors can be placed inside the loop of the basic resonator. In this way, the discrete resonators can be arranged and coupled in any desired lattice configuration. Furthermore, the platform is modular, in the sense that the basic LC-resonators can be fabricated independently in different shapes and configurations and then assembled into the targeted circuit via inductive couplings. Let us point out that, depending on the winding of the coupled inductors, the coupling parameter $M$ can be positive or negative and this is all that is needed to implement the entire classification table of topological insulators and superconductors \cite{ourpaper}.

\begin{figure*}
\center
  \includegraphics[width=0.7\linewidth]{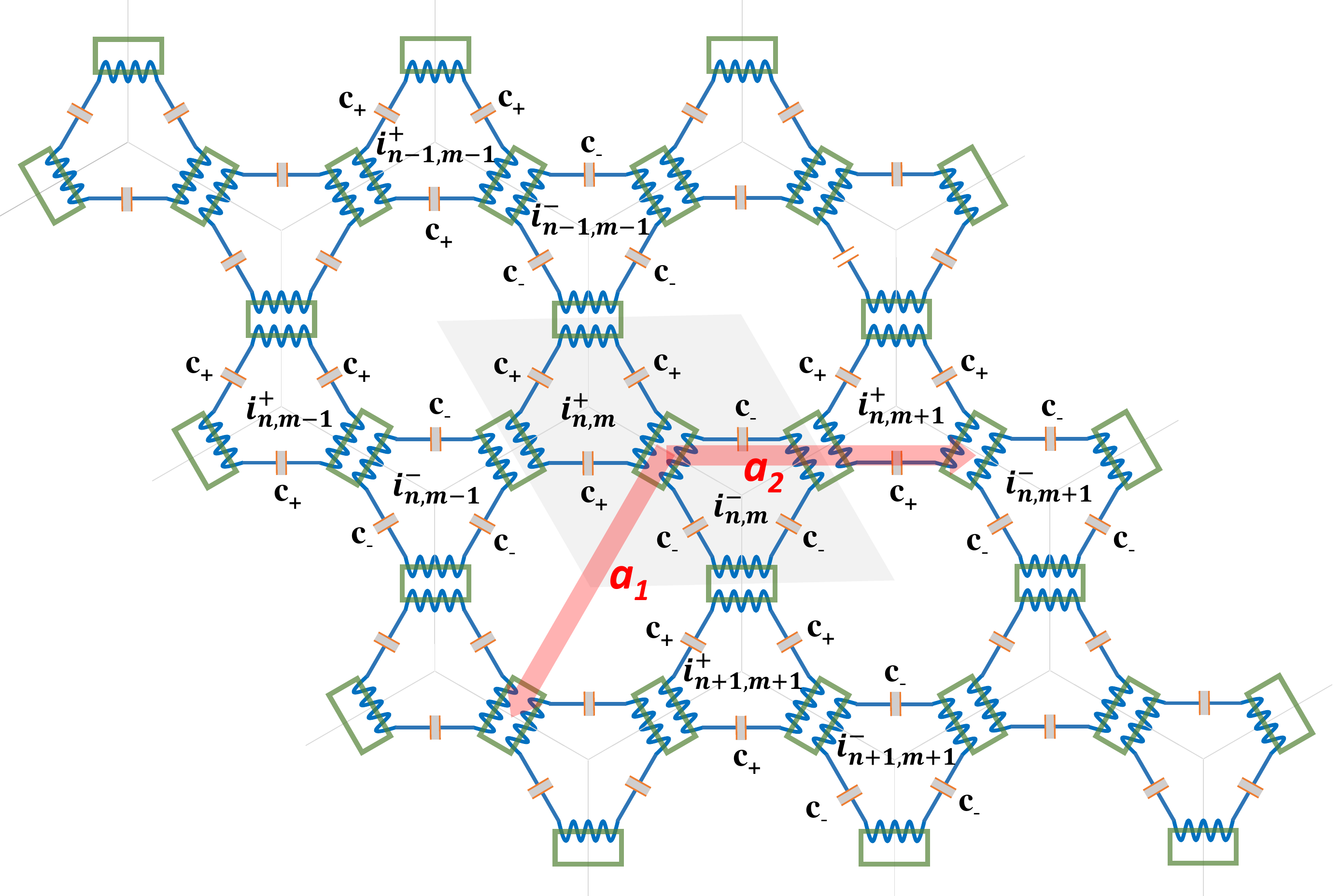}\\
  \caption{Honeycomb lattice of inductively coupled LC-resonators. Each vertex carries a closed LC-circuit, hence a conserving current, and the bonds carry the inductive couplings. The positive orientation of the currents is counter-clockwise. The diagram also shows a primitive cell (see the shaded region), the primitive vectors $\bm a_{1,2}$ (see the red thick arrows) and the labels of the vertices and currents. The values of the inductances and mutual-inductrances are uniform and fixed at $\frac{1}{3}L$ and $M$, respectively, while the capacitances take two values $c_\pm=\frac{1}{3}C_\pm$ in each primitive cell. }
 \label{Fig:CircuitLabeled}
\end{figure*}

\section{The Lattice of LC-Resonators}

We consider now an infinite lattice of coupled resonators as in Fig.~\ref{Fig:CircuitLabeled}. Each circuit loop sits atop of a vertex and and each inductive coupling sits atop of a link of a honeycomb lattice. As it is well known, the primitive cell of the honeycomb lattice contains two vertices (see shaded region in Fig.~\ref{Fig:CircuitLabeled}), which will carry the label $\alpha = \pm 1$. By shifting one reference cell by the primitive vectors $\bm a_1$ and $\bm a_2$, one can tile the entire plane. The primitive cells of this tilling can be labeled uniquely by two integers $(n,m)$. For example, the center of the $(n,m)$-primitive cell is located at $\bm R_{n,m} = n \bm a_1 + m\bm a_2$. As such, each vertex of the honeycomb lattice can be uniquely labeled by the three indices $(n,m,\alpha)$. 

The standard procedure to generate the valley-Chern effect is to break the inversion symmetry of the honeycomb lattice. We will achieve this here by setting different values to the total capacitance $C_\alpha$ for the two vertices of the primitive cells, while keeping $L$ and $M$ $\alpha$-independent. We should point out this is the most reasonable and practical choice that leads to the valley-Chern effect. For example, the effect will also appear if we introduce an $\alpha$-dependence on the inductances and keep the rest of the parameters uniform throughout the lattice. However, $L$ and $M$ are usually related, hence keeping $M$ uniform will require additional engineering.

Starting from Eqs.~\eqref{Eq:Coupling2} and \eqref{Eq:Coupling3} and guided by the labels supplied in Fig.~\ref{Fig:CircuitLabeled}, we derived the governing system of equations of the circuit, which take the form:
\begin{equation}\label{Eq:Partial1}
\bigg (\frac{1}{j \omega C_{\alpha}} + j \omega L \bigg ) \, i_{n,m}^\alpha + j \omega M \big ( i_{n,m}^{-\alpha} + i_{n-\alpha,m-\alpha}^{-\alpha} + i_{n,m-\alpha}^{-\alpha} \big ) = 0,
\end{equation}
for all $(n,m) \in \ZM^2$ and $\alpha=\pm 1$. In the absence of driving sources, these equations describe self-sustained oscillating currents, which can exist in the idealized scenario where the resistance of the loops is zero. The effect of the dissipative components will be discussed in a separate section.

\begin{figure*}
\center
  \includegraphics[width=0.7\linewidth]{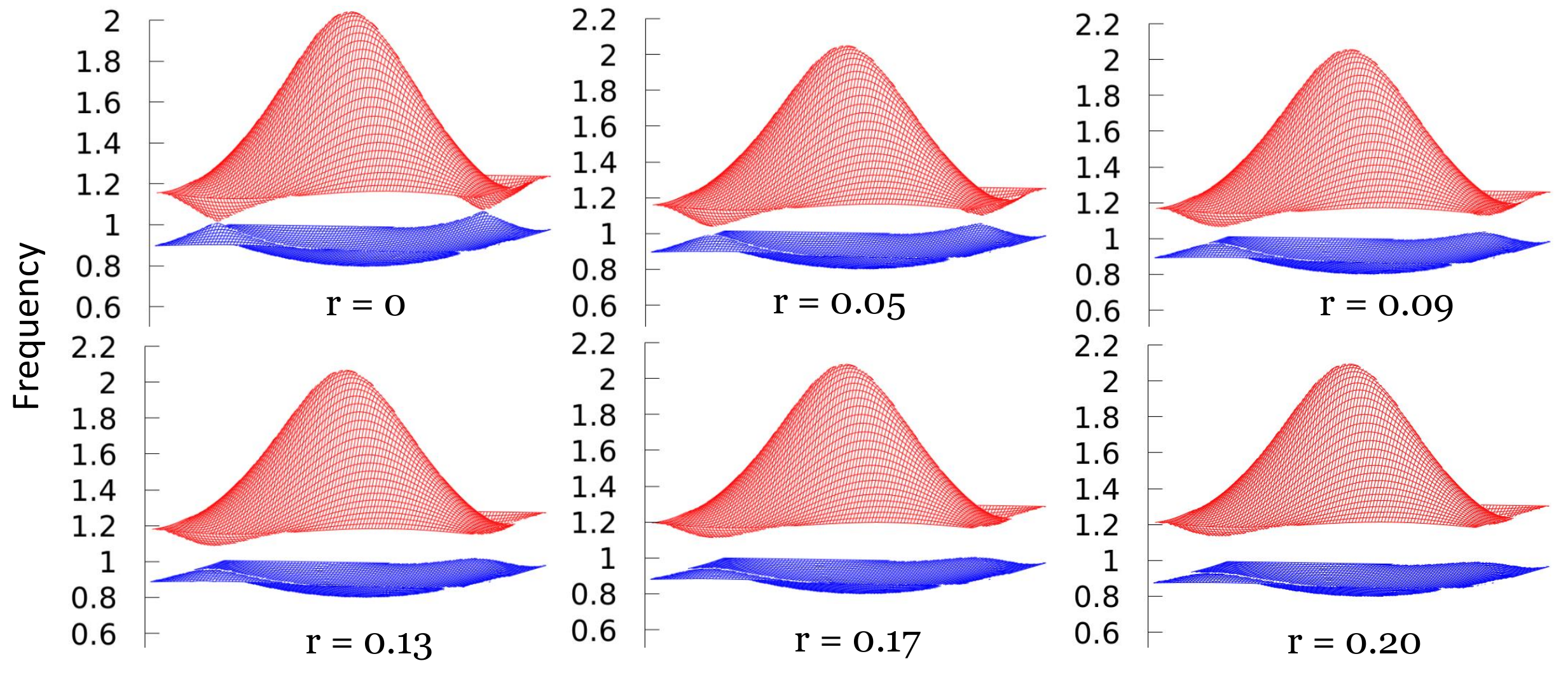}\\
  \caption{\small Dispersion surfaces of the bulk resonant modes as computed with Eq.~\eqref{Eq:BulkDispersion} for various values of $r$ and $\beta=0.25$. The graphs are rendered as functions of $(k_1,k_2)\in [-\pi,\pi]^2$ and the frequencies are expressed in units of $\nu_0$.}
 \label{Fig:TheoreticalBulk}
\end{figure*}

Our final goal for this section is to transform this system of equations into an eigenvalue problem defined on a suitable Hilbert space. After dividing Eq.~\eqref{Eq:Partial1} by $j\omega L$ and by introducing the dimensionless parameter $\beta = M/L$, we find:
\begin{equation}\label{Eq:Partial1}
\bigg(1- \frac{\omega^2_\alpha}{\omega^2} \bigg) \, i_{n,m}^\alpha +  \beta \, \big (i_{n,m}^{-\alpha} + i_{n-\alpha,m-\alpha}^{-\alpha} + i_{n,m-\alpha}^{-\alpha} \big )= 0,
\end{equation}
where $\omega_\alpha = 1/\sqrt{L C_\alpha}$ are the resonant pulsations of the decoupled resonators. At this point, we consider the following change of variables:
\begin{equation}
i^{\alpha}_{n,m} = \frac{1}{\omega_\alpha} q^\alpha_{n,m} \, , \quad (n,m) \in \ZM^2, \quad \alpha=\pm 1, 
\end{equation}
which transforms the equations into:
\begin{equation} 
\frac{1}{\omega^2} q_{n,m}^\alpha=\frac{1}{\omega_\alpha^2}q_{n,m}^\alpha +  \frac{\beta}{\omega_- \omega_+} (q_{n,m}^{-\alpha} + q_{n-\alpha,m-\alpha}^{-\alpha} + q_{n,m-\alpha}^{-\alpha}).
\label{eq:fundamental2}
\end{equation}
Together with the parameters:
\begin{equation}\label{Eq:Omega0}
\frac{1}{\omega_0^2} = \tfrac{1}{2} \bigg (\frac{1}{\omega_+^2}+\frac{1}{\omega_-^2} \bigg )=\tfrac{L}{2}\big (C_+ + C_- \big ),
\end{equation}
\begin{equation}
 r=\frac{\omega_+^{-2} -\omega_-^{-2}}{\omega_-^{-2} +\omega_+^{-2}} = \frac{C_+ - C_-}{C_+ + C_-},
\end{equation}
\begin{equation}
 \beta_r=\frac{\beta \omega_0^2}{\omega_- \omega_+}=\beta \sqrt{1-r^2},
\end{equation}
we can write the equations in the following form:
\begin{equation}\label{Eq:Dispersion0} 
\bigg ( \tfrac{\omega_0^2}{\omega^2} -1\bigg ) q_{n,m}^\alpha= \alpha r \, q_{n,m}^\alpha +  \beta_r (q_{n,m}^{-\alpha} + q_{n-\alpha,m-\alpha}^{-\alpha} + q_{n,m-\alpha}^{-\alpha}).
\end{equation}

We collect all the data in a function $Q: \mathbb Z^2 \rightarrow \mathbb C^2$:\begin{equation}
Q(n,m)= 
\begin{pmatrix} 
q_{n,m}^{+1} \\ q_{n,m}^{-1} \end{pmatrix} \in \CM^2,
\end{equation}
and, on the space of these functions, we introduce the scalar product:
\begin{equation}\label{Eq:ScalarProduct}
\langle Q,Q'\rangle = \sum_{(n,m)\in \ZM^2} Q(n,m)^\dagger Q'(n,m).
\end{equation}
This supplies the Hilbert space $\ell^2(\ZM^2,\CM^2)$ of square-summable sequences over the lattice with values in $\CM^2$. In Eq.~\eqref{Eq:ScalarProduct}, the dagger stands for $\begin{pmatrix} \xi_1 \\ \xi_2 \end{pmatrix}^\dagger = (\xi_1^\ast,\xi_2^\ast)$. Furthermore, we introduce the shift operators:
\begin{align}
& (S_1 Q)(n,m) = Q(n+1,m), \\ 
& (S_2 Q)(n,m)=Q(n,m+1),
\end{align}
which enable us to write the dispersion equations in the following compact form: 
\begin{equation}\label{Eq:Dispertion1}
\begin{array}{c}
\bigg (\frac{\omega_0^2}{\omega^2}- 1 \bigg ) Q =   
\\ \nonumber 
  \left [  r 
\begin{pmatrix} 
1 & 0 \\
0 & -1
\end{pmatrix} 
  +  \beta_r \begin{pmatrix} 
0 & 1+S_2^\dagger S_1^\dagger +S_2^\dagger 
\\ 1+S_1 S_2 + S_2 & 0 
\end{pmatrix} 
 \right] Q.
 \end{array}
\end{equation}

\begin{figure*}
\center
 \includegraphics[width=0.6\linewidth]{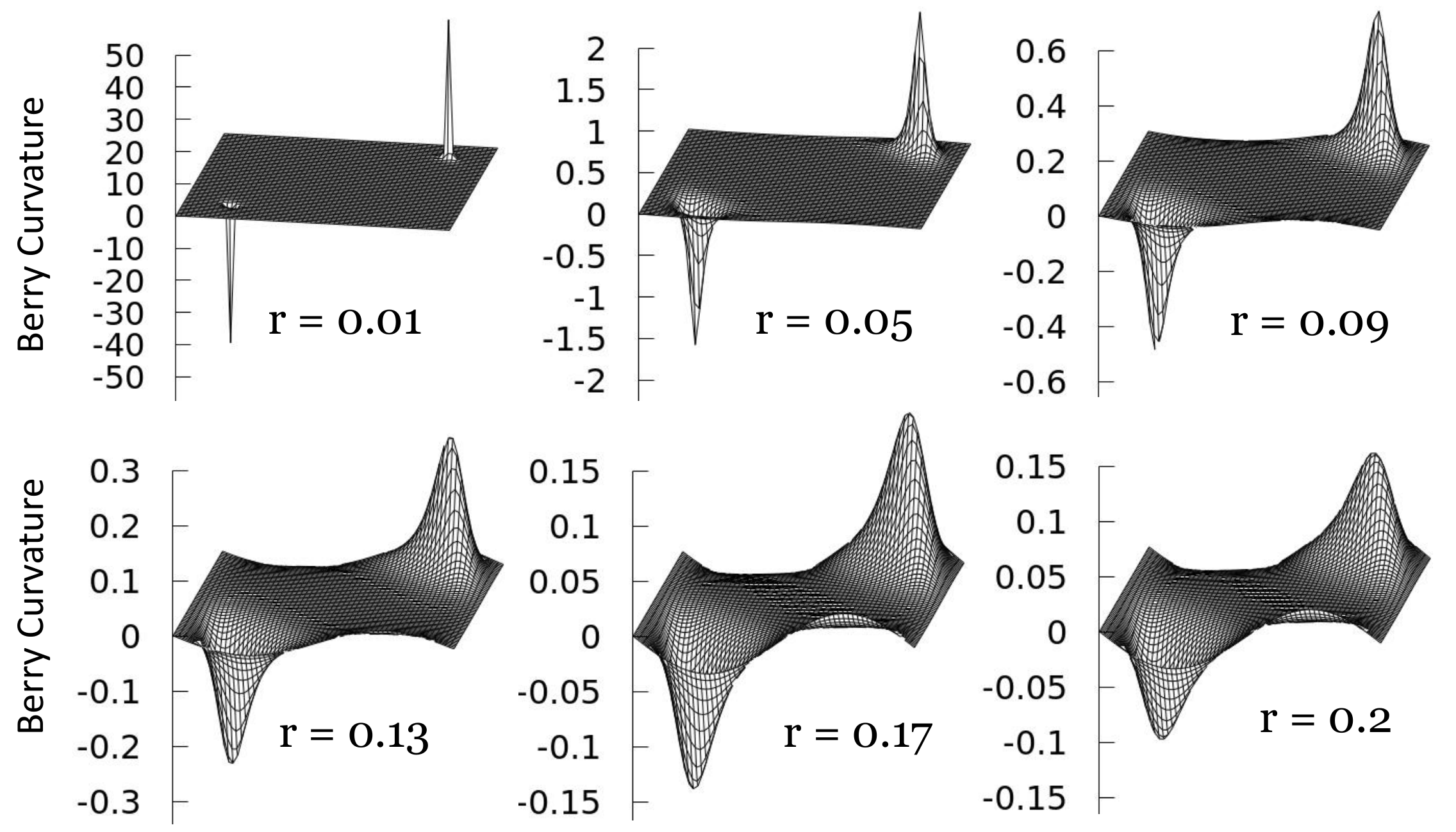}\\
  \caption{\small Maps of the Berry curvature for various values of $r$ and $\beta=0.25$, as a function of  $(k_1,k_2)\in [-\pi,\pi]^2$.}
 \label{Fig:BerryPlots}
\end{figure*}

As promised, we have reduced the problem of finding the resonant modes of the circuit to the eigenvalue problem $\lambda Q = D Q$ defined on the Hilbert space $\ell^2(\ZM^2,\CM^2)$, with $\lambda = \omega_0^2/\omega^2 -1$ and:
\begin{equation}\label{Eq:RSDynMat}
D=r \sigma_3 
+ \beta_r \big ( \sigma_1+ (S_1+1)S_2 \sigma_- + S_2^\dagger (S_1^\dagger   + 1)\sigma_+ \big ),
\end{equation}
where $\sigma$'s are Pauli's matrices.

\section{Dispersion of the Bulk Modes}
\label{Sec:BulkDispersion}

In this section, we collect the two spatial indices $(n,m)$ into one label $\bm n \in \ZM^2$. Let us consider the following functions:
\begin{equation}\label{Eq-PlaneWave}
\phi_{\bm k}(\bm n) = e^{\imath \bm k \cdot \bm n}, \quad \bm k \in \TM^2,
\end{equation}
where $\TM^2$ is the flat 2-torus. It is straightforward to check that $\phi_{\bm k}$'s are common eigenvectors of the shift operators: 
\begin{equation}
\big(S_j \phi_{\bm k}\big )(\bm n) = e^{\imath k_j} \phi_{\bm k}(\bm n).
\end{equation} 
As such, we seek the eigen-modes of $D$ in the form $\bm \xi \phi_{\bm k}$, $\bm \xi \in \CM^2$, because then $D (\bm \xi \phi_{\bm k} ) = \big (\widehat D(\bm k)\bm \xi\big ) \phi_{\bm k}$, with $\widehat D (\bm k)$ the $2 \times 2$ $\bm k$-dependent matrix:
\begin{equation}\label{Eq:DK}
\widehat D (\bm k) = r \sigma_3 
+ \beta_r \big (f(\bm k)\sigma_- +f(\bm k)^\ast \sigma_+  \big ),
\end{equation}
where:
\begin{equation}
f(\bm k) = 1 + e^{\imath k_2} + e^{\imath (k_1+k_2)}.
\end{equation}
The spectrum of the dynamical matrix can now be computed explicitly:
\begin{equation}\label{Eq:Lambda}
\lambda_\pm (\bm k) = \pm \sqrt{r^2 +\beta_r^2|f(\bm k)|^2},
\end{equation}
and this leads to the resonant frequencies:
\begin{equation}\label{Eq:BulkDispersion}
\nu_\pm(\bm k) = \frac{\nu_0}{\bigg [1\mp \sqrt{r^2 +\beta_r^2|f(\bm k)|^2}\, \bigg ]^\frac{1}{2}},
\end{equation}
where $\nu_0$ is the frequency corresponding to the pulsation $\omega_0$ defined in Eq.~\eqref{Eq:Omega0}.

A graphic representation of these dispersion equations are reported in Fig.~\ref{Fig:TheoreticalBulk}, for several values of the parameter $r$. For $r=0$ one can see the two dispersive bands touching at two points, where conical singularities occur. These are the Dirac cones, well known from the physics of graphene, which occur at the following two particular $k$-points:
\begin{equation}
\bm K =-\bm K'= \big (\tfrac{2 \pi}{3},-\tfrac{2\pi}{3} \big ).
\end{equation} 
The dispersion equations for $\bm k \simeq \bm K$ or $\bm k \simeq \bm K'$ is well captured by an effective mass-less Dirac Hamiltonian \cite{RenRPP2016}. As soon as $r$ is turned on, the Dirac cones split and a spectral gap emerges. Examining the upper dispersion band, one sees the defunct Dirac singularities appearing as deep valleys in the dispersion landscape, and this explains why $K$ and $K'$ are called valleys. For small values of $r$ and near the valleys, the dispersion surfaces are well characterized by a massive Dirac effective Hamiltonian \cite{RenRPP2016}. As it is well known, these Dirac effective models can be used to understand the bulk-boundary correspondence in QVHE. Note, however, that the valleys become shallower as the parameter $r$ is increased, and that the region where the effective Dirac models can be applied shrinks rapidly. As such, even though the bulk gap is enhanced when increasing $r$, QVHE is expected to become weaker. This aspect will be further substantiated by the analysis of the Berry curvature and of the interface modes.

\begin{figure*}
\center
  \includegraphics[width=0.7\linewidth]{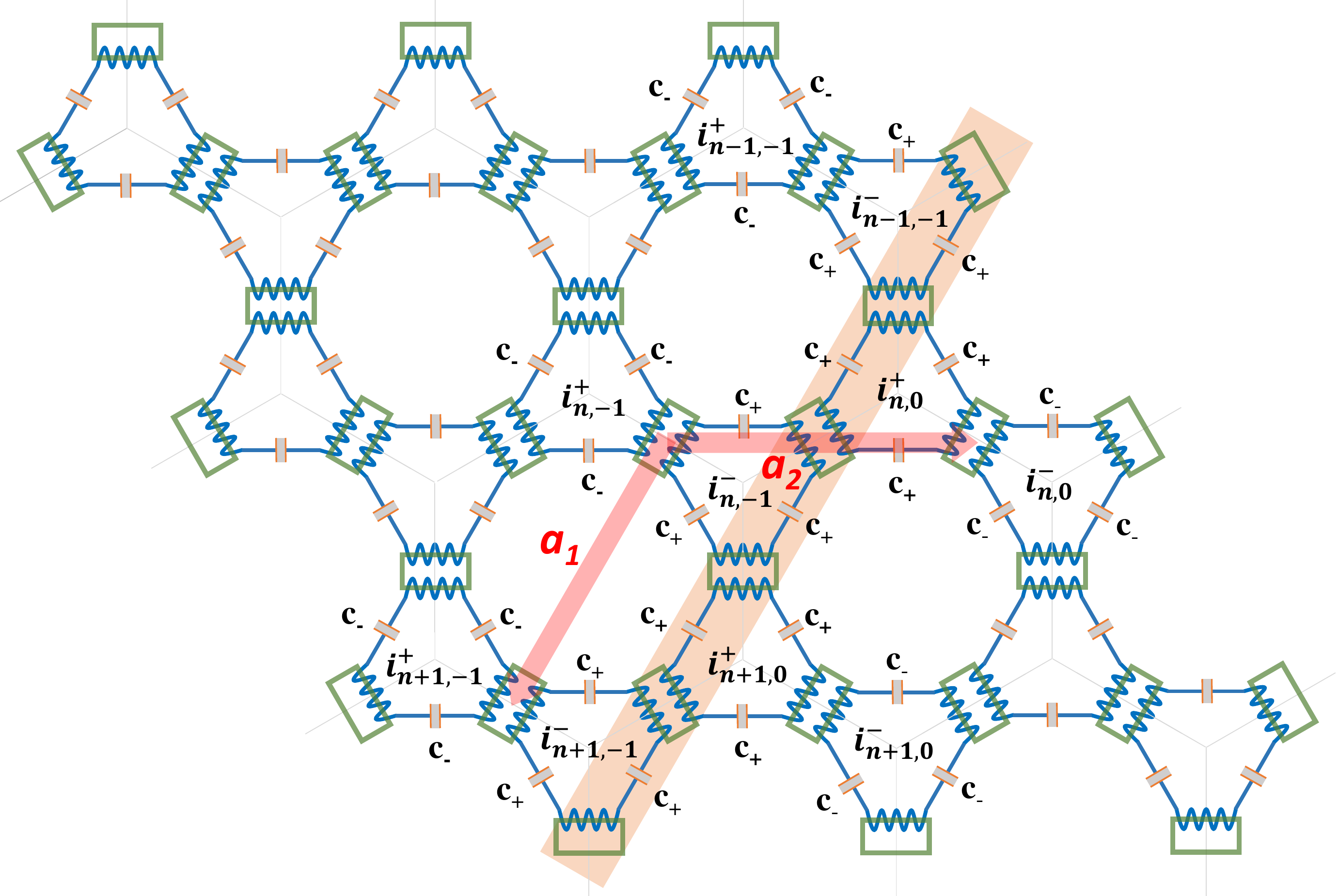}\\
  \caption{\small The configuration with a domain wall. The domain wall, highlighted by the shaded region along the primitive vector $\bm a_1$, joins two reflection-inverted honeycomb lattices of LC-resonators. The labels of the primitive cells remain as in Fig.~\ref{Fig:CircuitLabeled} and only the values of the capacitors are changed across the domain wall.}
 \label{Fig:DomainWall}
\end{figure*}

\subsection{Analysis of the Berry Curvature}
\label{Sec:Berry}

The landscape of the Berry curvature is essential for understanding the topological nature of the valley-Chern effect. Indeed, the emergence of quasi-chiral interface modes along a domain wall has its topological origin in the concentration of the Berry curvature near the valleys \cite{QianPRB2018}. These aspects will be addressed in the following section. Here, we map the Berry curvature and assess the parameter values for which the valley-Chern effect is expected to be strong.

For gapped dispersion surfaces, the gap projection is defined as:
\begin{equation}\label{Eq:GapProjection1}
P_G(\bm k) = \chi_{[-\infty,G]}\big ( D(\bm k) \big ),
\end{equation}
where $\chi$ is the indicator function and $G$ is the mid-gap frequency. The right-hand side of Eq.~\eqref{Eq:GapProjection1} is a function evaluated on a Hermitean matrix, which can be computed by either appealing to the spectral decomposition of the matrix or using polynomial approximations of the function. For a model with only two dispersion surfaces, the gap projection can be computed as:
\begin{equation}\label{Eq:Pg1}
\widehat P_G(\bm k) =\frac{\widehat D(\bm k) - \lambda_+(\bm k)}{\lambda_{-}(\bm k) - \lambda_{+}(\bm k)}.
\end{equation} 
Things become even more manageable when the dynamical matrix takes the form $D(\bm k) = \bm v(\bm k) \cdot \bm \sigma$, as in our case. The explicit expression of $\bm v$ can be derived from Eq.~\ref{Eq:DK}:
\begin{equation}
\bm v(\bm k) = \big ( \beta_r {\rm Re}[f(\bm k)], \beta_r {\rm Im}[f(\bm k)], r\big ).
\end{equation}
In this case:
\begin{equation}\label{Eq:Pg2}
P_G (\bm k) = \tfrac{1}{2}(I - \hat{\bm v}(\bm k) \cdot {\bm \sigma}), \quad \hat {\bm v}(\bm k)=\bm v /|\bm v|.
\end{equation} 

Associated to $P_G$ is the Berry curvature \cite{AvronCMP1989}:
\begin{equation}\label{Eq:Curvature}
F(\bm k) = (2 \pi \imath)^{-1} \, {\rm Tr} \big( P_G(\bm k) \big[ \partial_{k_{1}} P_G (\bm k),\partial_{k_{2}} P_G (\bm k) \big] \big),
\end{equation}
where ${\rm Tr}$ is the trace over the two internal degrees of freedom. Using Eq.~\eqref{Eq:Pg2}, together with elementary properties of Pauli's matrices, the above expression can be evaluated to:
\begin{equation}\label{Eq:BerryFormula1} 
F(\bm k) = \frac{1}{4\pi} \,  \hat {\bm v} \cdot (\partial_{k_{1}} \hat{\bm v} \times \partial_{k_{2}} \hat{\bm v}).
\end{equation}
For our specific system, we find: 
\begin{equation}\label{Eq:BerryExpression}
F(\bm k) = \frac{1}{4\pi} \frac{r\beta_r^2 \sin(k_1)}{ \big(r^2+ \beta_r^2 |f(\bm k)|^2 \big)^{3/2}}.
\end{equation}

A graphical representation is reported in Fig.~\ref{Fig:BerryPlots} for various values of the parameter $r$. As expected for systems with time-reversal symmetry, the Berry curvature is odd under the inversion: $F(\bm k) = - \bm F(\bm k)$. Among other things, this implies that the total Chern number of the dispersion band vanishes. As discovered in \cite{QianPRB2018}, the topological character of the valley-Chern effect steams from the concentration of the Berry curvature near the valleys. For example, in the limit $r \searrow 0$, it is known that the Berry curvature converges to the singular distribution $\frac{1}{2}\delta(\bm k - \bm K) - \frac{1}{2} \delta (\bm k - \bm K')$. This is well reflected in the distributions corresponding to the lowest value $r=0.01$ shown in Fig.~\ref{Fig:BerryPlots}. As $r$ is increased, the distribution of the Berry curvature  broadens yet in the three upper panels of \ref{Fig:BerryPlots}, it still remains concentrated near the valleys. In these cases, the valley-Chern effect is expected to be strong. For the lower panels, however, the Berry curvature is not only broadened but a significant part of it has been lost. To quantify the latter statement, we compute the integral of the Berry curvature over half of the Brillouin zone $k_1+k_2\leq 0$ and the result are 0.48, 0.41, 0.34, 0.29, 0.24, 0.21 for $r=0.01$, $0.05$, $0.09$, $0.13$, $0.17$, $0.20$, respectively. These numbers illustrate the gradual diminishing of the Berry curvature supported by each valley as $r$ is increased, which actually correlates with the flattening of the dispersion surfaces near the valleys, observed in the previous section.

The conclusion is that the system looses the engine of the topological protection as $r$ is increased. As such, one needs to compromise between the size of the bulk gap, which determines the localization of the DW-modes along the interfaces, and the topological protection of the modes. As reported in \cite{QianPRB2018}, this is a limitation of the simple first near-neighbor implementation of the valley-Chern effect, which can be, in principle, corrected by band and Berry curvature engineering using further near-neighbor couplings. As we shall see, all these observations have important physical consequences on the domain wall modes.

\begin{figure*}
\center
  \includegraphics[width=0.7\linewidth]{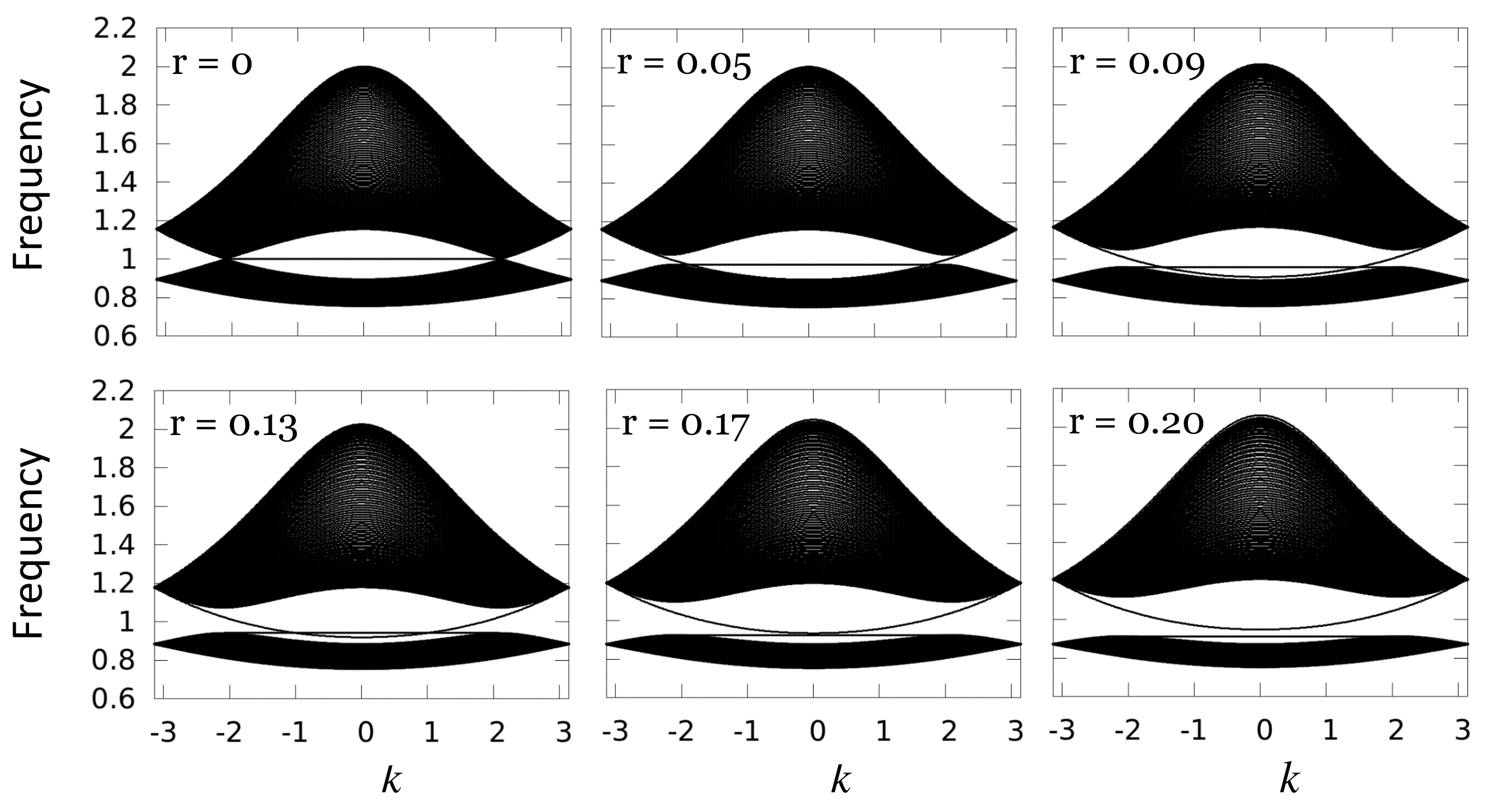}\\
  \caption{\small The predicted domain wall modes for various values of $r$ and $\beta=0.25$. The data was generated with Eq.~\ref{Eq:DW5} with $m$ restricted in the finite interval $[-50,50]$. Dirichlet boundary conditions were imposed at the end of this interval, leading to the flat bands seen in each of the panels. Since the domain wall is in the middle of the interval, these flat bands have nothing to do with the physics investigated here.}
 \label{Fig:TheoreticalEdge}
\end{figure*}

\section{Configuration with a Domain-Wall}
\label{Sec:DW}

The circuit with a domain wall is illustrated in Fig.~\ref{Fig:DomainWall}. In this diagram, one can see an interface along $\bm a_1$, located between $m =1$ and $m=0$. Let us specify, explicitly, that the labels attached to the unit cells are not modified but only the values of the capacitors. The system displays a reflection symmetry relative to this interface, with the reflection operator acting as: 
\begin{equation}
\mathcal I |n_1,n_2,\alpha \rangle = |n_1,- n_2-1, -\alpha \rangle.
\end{equation}
Note that the couplings between the individual resonators remain the same.

\begin{figure}[b]
\center
  \includegraphics[width=\linewidth]{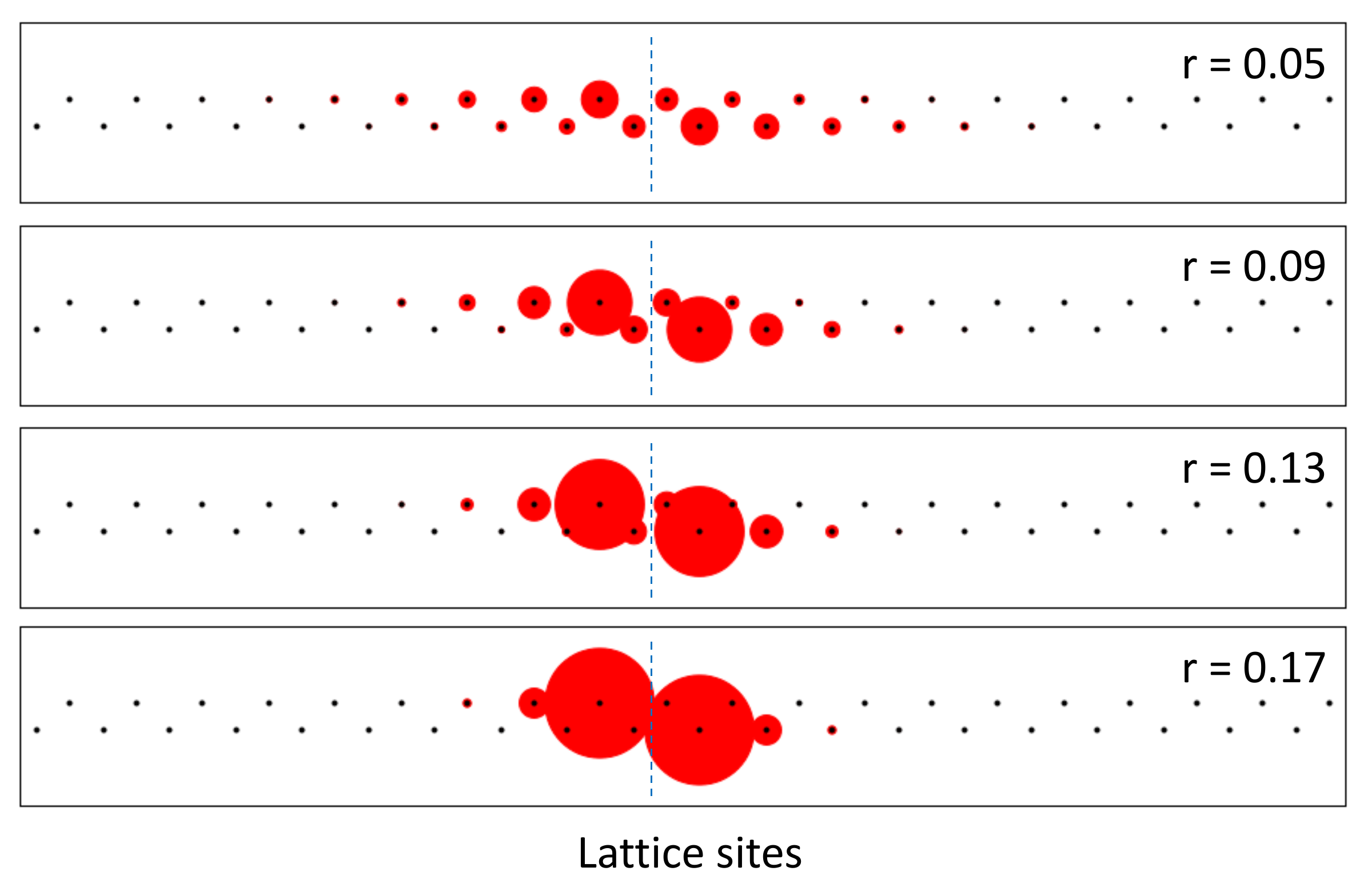}\\
  \caption{\small Localization of the edge modes for various values of $r$ and $\beta=0.25$. The dash vertical line indicates the position of the interface. Since the translation symmetry is preserved along $\bm a_1$, only a row of primitive cells along $\bm a_2$ is shown. The amplitude of the DW-mode at frequency $\nu_0$, located in the middle of the bulk spectral gap, is proportional to the size of the red disks.}
 \label{Fig:EdgeModes}
\end{figure}

In the so-called continuum limit, which applies when the bulk spectral gaps are small, the bulk-boundary correspondence for the valley-Chern effect can be entirely understood from the effective Dirac models. Indeed, the decoupled Dirac models corresponding to the two valleys can be transformed from quasi-momentum to real-space coordinates, in which case the effect of a domain-wall can be investigated. The result is well known \cite{XiaoPRL2007}: chiral modes localized near the interface emerge from each valley. An alternative approach was proposed in \cite{QianPRB2018}, where it was observed that the system with a domain-wall can be folded and transformed into a bi-layered system with an edge. The difference is that, now, each valleys carries an integer quanta of Berry curvature and, hence there are no topological obstruction for continuing each of the valleys to full separate bands. Then the valleys become band indices and the valley-Chern effect can be rigorously connected to the spin-Chern effect. The chiral interface modes mentioned above can then be understood as the standard chiral edge associated to spin-Chern insulators \cite{ShengPRL2006,ProdanPRB2009}.

In this section, we will witness this phenomenon in the honeycomb lattice of LC-resonators. Using the labels supplied in Fig.~\ref{Fig:DomainWall}, in the presence of a domain wall the dispersion equations \eqref{Eq:Partial1}
 become:
\begin{equation}\label{Eq:DW1}
\bigg(1- \tfrac{\omega^2_\alpha}{\omega^2} \gamma_m^\alpha\bigg) \, i_{n,m}^\alpha +  \beta \, \big (i_{n,m}^{-\alpha} + i_{n-\alpha,m-\alpha}^{-\alpha} + i_{n,m-\alpha}^{-\alpha} \big )= 0,
\end{equation}
where $\gamma_m^\alpha = 1$ for $m \geq 0$ and $\gamma_m^\alpha = \omega_{-\alpha}^2/\omega_\alpha^2$ otherwise. The latter can be written compactly as:
\begin{equation}
\gamma_m^\alpha = 1+\frac{2\alpha r}{1-\alpha r}\chi_{(-\infty,0)}(m),
\end{equation}
where $\chi$ is the indicator function.

Our first goal is to transform these coupled equations into a genuine eigen-problem on the Hilbert space $\ell^2(\ZM^2,\CM^2)$. For this, we perform the change of variable $i_{n,m}^\alpha \rightarrow q_{n,m}^\alpha$ as before and transform the equations into:
\begin{equation}\label{Eq:DW2} 
\tfrac{\omega_0^2}{\omega^2} \gamma^\alpha_m q_{n,m}^\alpha= (1+\alpha r) \, q_{n,m}^\alpha +  \beta_r (q_{n,m}^{-\alpha} + q_{n-\alpha,m-\alpha}^{-\alpha} + q_{n,m-\alpha}^{-\alpha}).
\end{equation}
Collecting the data into the function $Q$ and using the shift operators as well as the new diagonal operator:
\begin{equation}
\big ( \Gamma Q \big )(n,m) =  \begin{pmatrix} \gamma_m^+ & 0 \\ 0 & \gamma_m^- \end{pmatrix} \, Q(n,m),
\end{equation}
the above equations can be written compactly as:
\begin{align}\label{Eq:DW3}
& \qquad \quad \tfrac{\omega_0^2}{\omega^2} \Gamma Q =  (1+r\sigma_3)Q  
\\ \nonumber 
  & +  \beta_r \begin{pmatrix} 
0 & 1+S_2^\dagger S_1^\dagger +S_2^\dagger 
\\ 1+S_1 S_2 + S_2 & 0 
\end{pmatrix}  Q.
\end{align}

\begin{figure}[t]
\center
\includegraphics[width=\linewidth]{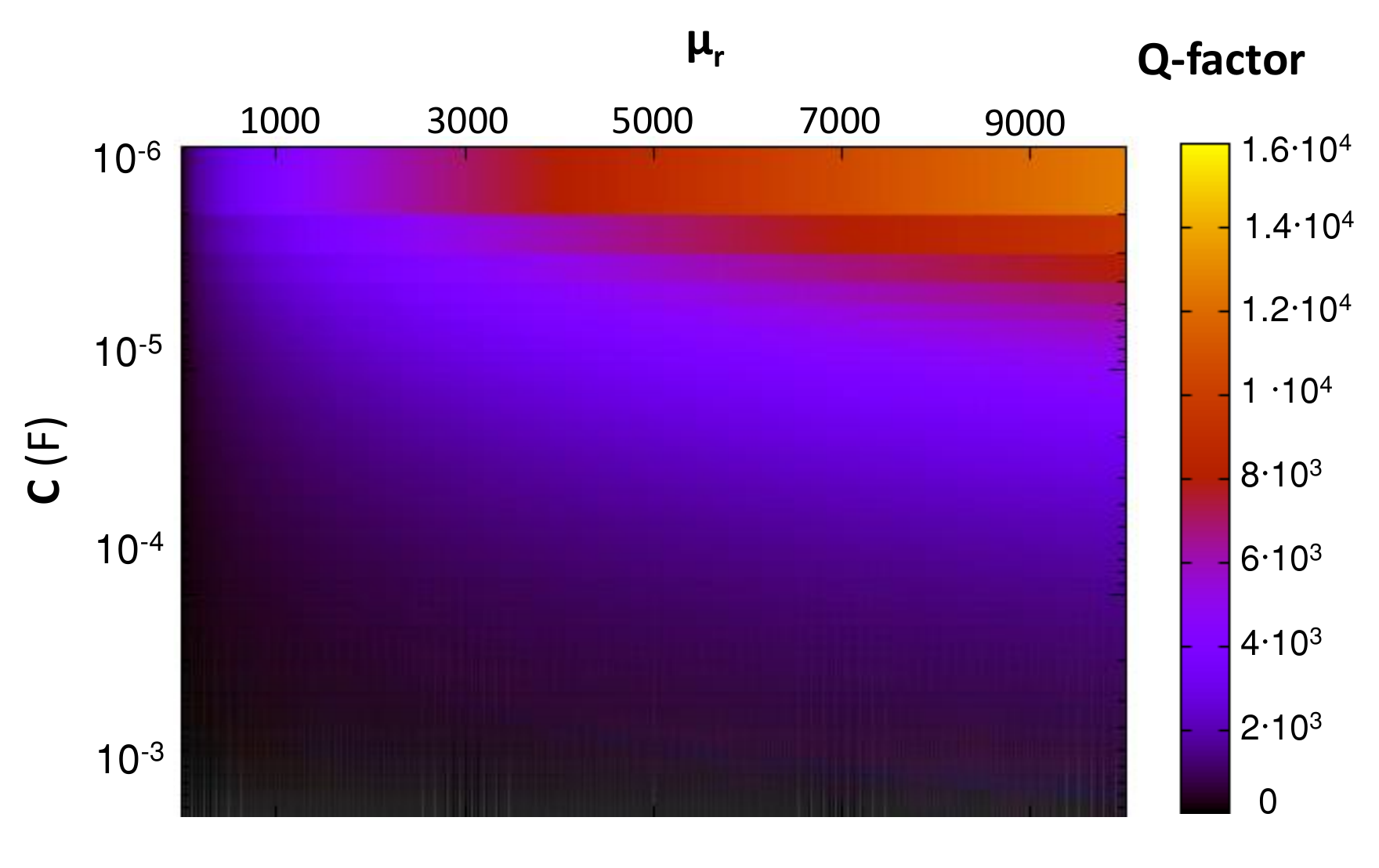}
  \caption{Map of the Q-factor as function of $c$ and $\mu_r$. The other parameters are fixed at: $d=1$ mm, $\rho= 1.68\times 10^{-8}$, $N=10$. The resonant frequency over the entire map lies inside the range $[101.3,\,320405.7]$~Hz.
}
 \label{Fig:QvsCandMur}
\end{figure}

The dispersion equation can now be transformed into a genuine eigen-value problem with the help of the following change of variable, $Q \rightarrow \Gamma^{\frac{1}{2}} Q'$, which leads to:
\begin{align}\label{Eq:DW4}
& \qquad \qquad \qquad \tfrac{\omega_0^2}{\omega^2} Q' = \Gamma^{-\frac{1}{2}}\left [ (1 + r\sigma_3)   \right .
\\ \nonumber 
 & \left . 
 +  \beta_r \big (1+S_2^\dagger S_1^\dagger +S_2^\dagger \big) \sigma_- +\big ( 
 1+S_1 S_2 + S_2 \big ) \sigma_+ 
 \right] \Gamma^{-\frac{1}{2}}Q',
\end{align}
Since $S_1$ commutes with the operator on the right, we can seek the eigen-modes in the form $Q'(n,m)=e^{\imath k n} Q'_k(m)$, with $Q'_k(m)$ an eigen-mode of:
\begin{align}\label{Eq:DW5}
& \qquad \tfrac{\omega_0^2}{\omega^2} Q'_k = \Gamma^{-\frac{1}{2}}\left [ (1 + r\sigma_3) +  \beta_r \sigma_1   \right .
\\ \nonumber 
 & \left . 
  +(1+e^{-\imath k})S_2^\dagger \sigma_- + (1+e^{\imath k})  S_2 \sigma_+ 
 \right] \Gamma^{-\frac{1}{2}}Q'_k.
\end{align}
The dispersion of the modes, as derived from \eqref{Eq:DW5}, is reported in Fig.~\ref{Fig:TheoreticalEdge}. There, one can see the bulk spectrum appearing as dark regions, which are just a side-view of the spectra reported in Fig.~\ref{Fig:TheoreticalBulk}. Focusing now on the valleys, let us point to the chiral bands seen to traverse the bulk spectral gap. These are the topological domain-wall modes. Note, however, that, as $r$ is increased, the chiral character weakens and at $r=0.2$ a spectral gap opens and the system is no longer metallic. The cause of this has been already anticipated in section~\ref{Sec:Berry} as the dilution of the Berry curvature supported by each valley. It is then apparent that, in this simple implementation of the valley-Chern effect, one is constraint to consider the lower values of $r$ parameter. However, as anticipated in section~\ref{Sec:BulkDispersion}, the bulk spectral gaps become small leading to a possible delocalization of the DW-modes.

To investigate the latter issue, we plot in Fig.~\ref{Fig:EdgeModes} the DW-modes as computed from \eqref{Eq:DW5}. As anticipated, for a value as low as $r=0.05$, the DW-mode is highly delocalized, while for $r=0.17$ the DW-modes are highly localized along the interface. Note that $r=0.17$ is the critical value where the edge spectrum becomes gapped. Finally, weighting both Fig.~\ref{Fig:TheoreticalEdge} and Fig.~\ref{Fig:EdgeModes}, we conclude that the optimal value of the parameter is $r=0.13$. At this value $r=0.13$, the back-scattering between the left and right-moving wave guiding modes is minimized and these modes are well-localized, resulting in both strong localization and well defined chirality. 

\begin{figure}[t]
\center
\includegraphics[width=\linewidth]{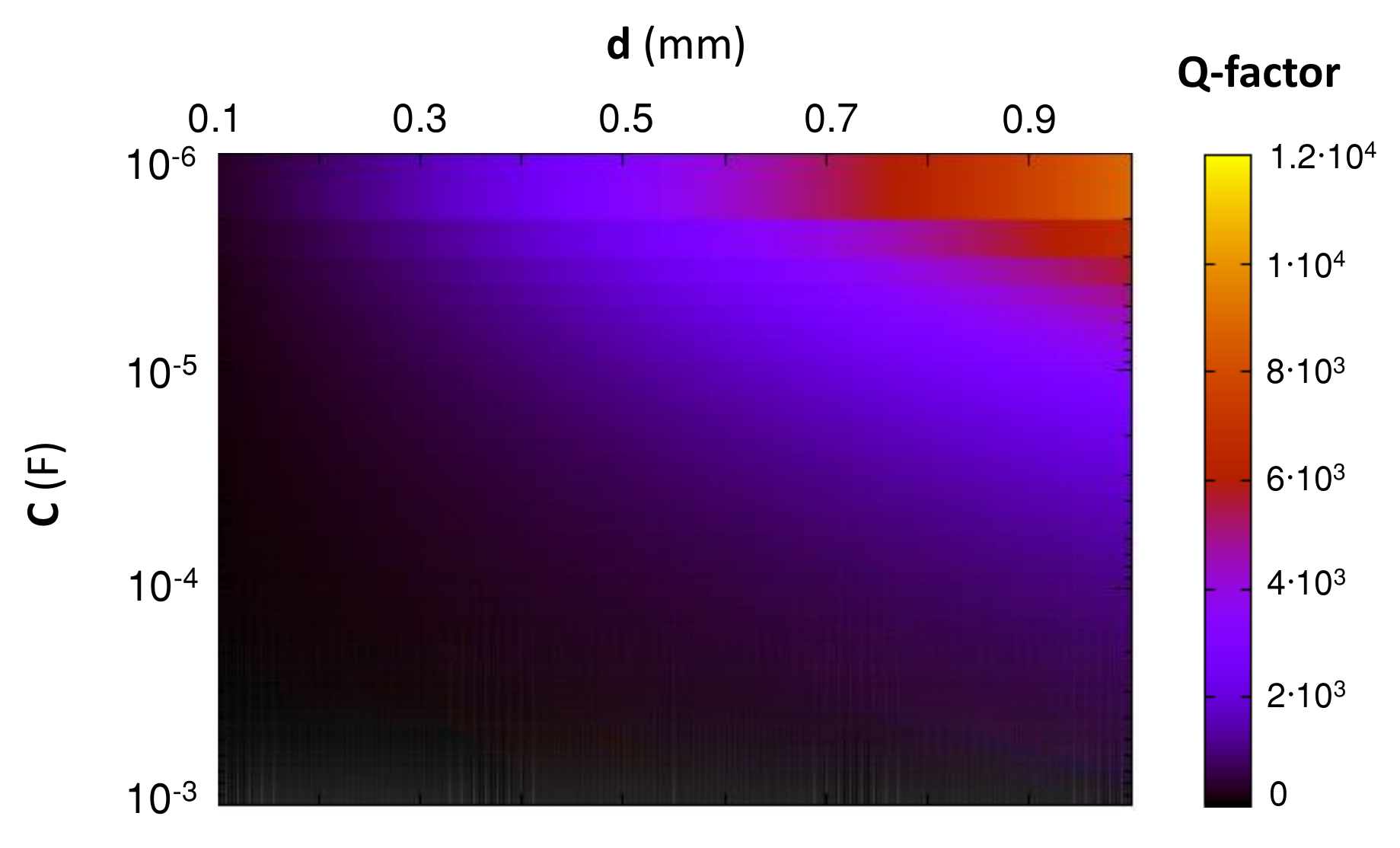}
  \caption{Map of the Q-factor as function of $c$ and $d$. The other parameters are fixed at: $\mu_r=5000$ m, $\rho= 1.68\times 10^{-8}$, $N=10$. The resonant frequency over the entire map lies in the range $[45.3,\,4531.2]$~Hz.}
 \label{Fig:QvsCandD}
\end{figure}

\section{Q-Factors and Experimental Parameters}

In this section we analyze a circuit built with classical solenoids. Our goal is to demonstrate the extreme high Q-factors that can be obtained with such a simple setup.

Henceforth, consider our basic resonator where, this time, we will take into account the resitance $R$ of the wire making up the inductor. The total impedance of the resonator is: 
\begin{equation}
Z=\frac{1}{j \omega_{0} C}+j\omega_{0} L +R,
\end{equation}
leading to a complex pulsation:
\begin{equation}
\omega_{0}=\sqrt{\frac{1}{LC} - \frac{R^2}{4L^2}}+j  \frac{R}{2L} \approx \frac{1}{\sqrt{LC}} + j  \frac{R}{2L}.
\end{equation}
The $Q$-factor of the LC-resonator is:
\begin{equation}
Q \approx \frac{\textrm{Re[}\omega_0 \textrm{]}}{2\textrm{Im[}\omega_0 \textrm{]}}=\frac{1}{R}\sqrt{\frac{L}{C}}.
\end{equation}

Now, consider that each solenoid in Fig.~\ref{Fig:Coupling} has length $\ell$, diameter $D$ and number of turns $N$. Furthermore, let the wire in each solenoid have diameter $d$ and total length $L_{s}$. Then: 
\begin{equation}
L = 3\frac{\pi \mu_0 \mu_r N^2 D^2}{4 \ell}, \quad R=3 \frac{4\rho L_{s}}{\pi d^{2}}, 
\end{equation} 
where $\mu_0$ is the magnetic permeability of the vacuum, $\mu_r$ is the relative permeability of the ferromagnetic bar and $\rho$ is the electrical resistivity of the wire. The latter will be fixed at $\rho = 1.68\times 10^{-8}$ $\Omega\cdot$m, appropriate for copper. Taking into account the inter-dependencies between the parameters:
\begin{equation}
\ell \approx Nd, \quad L_{s}=\pi D N,
\end{equation}
and that $C=\frac{1}{3}c$, we obtain:
\begin{equation}
Q=\sqrt{\frac{L}{CR^2}}= \sqrt{\frac{9\pi \mu_0 \mu_r N^2 D^2}{4 Ndc}\bigg (\frac{\pi d^2}{12 \rho \pi N D}\bigg)^2}
\end{equation}
or:
\begin{equation}
Q=\frac{1}{8} \sqrt{\frac{\pi \mu_0 \mu_r d^3}{\rho^2Nc}}.
\end{equation}

A map of the Q-factor as function of $c$ and $\mu_r$ is reported in Fig.~\ref{Fig:QvsCandMur}, with the other parameters fixed at $N=10$ and $d=1$~mm. Note that with this values, the length of the solenoid will be $\ell = 1$ cm.  From this data, we can see that for an iron bar with $\mu_r=5000$ we can generate Q-factors as high as $8\times 10^3$, while for the special materials with $\mu_r \sim 10^4$, such as cobalt-iron, the $Q$-factor can be larger than $10^4$. Note that this high Q-factors are obtained at the higher end of the frequency range.

\begin{figure}[t]
\center
\includegraphics[width=\linewidth]{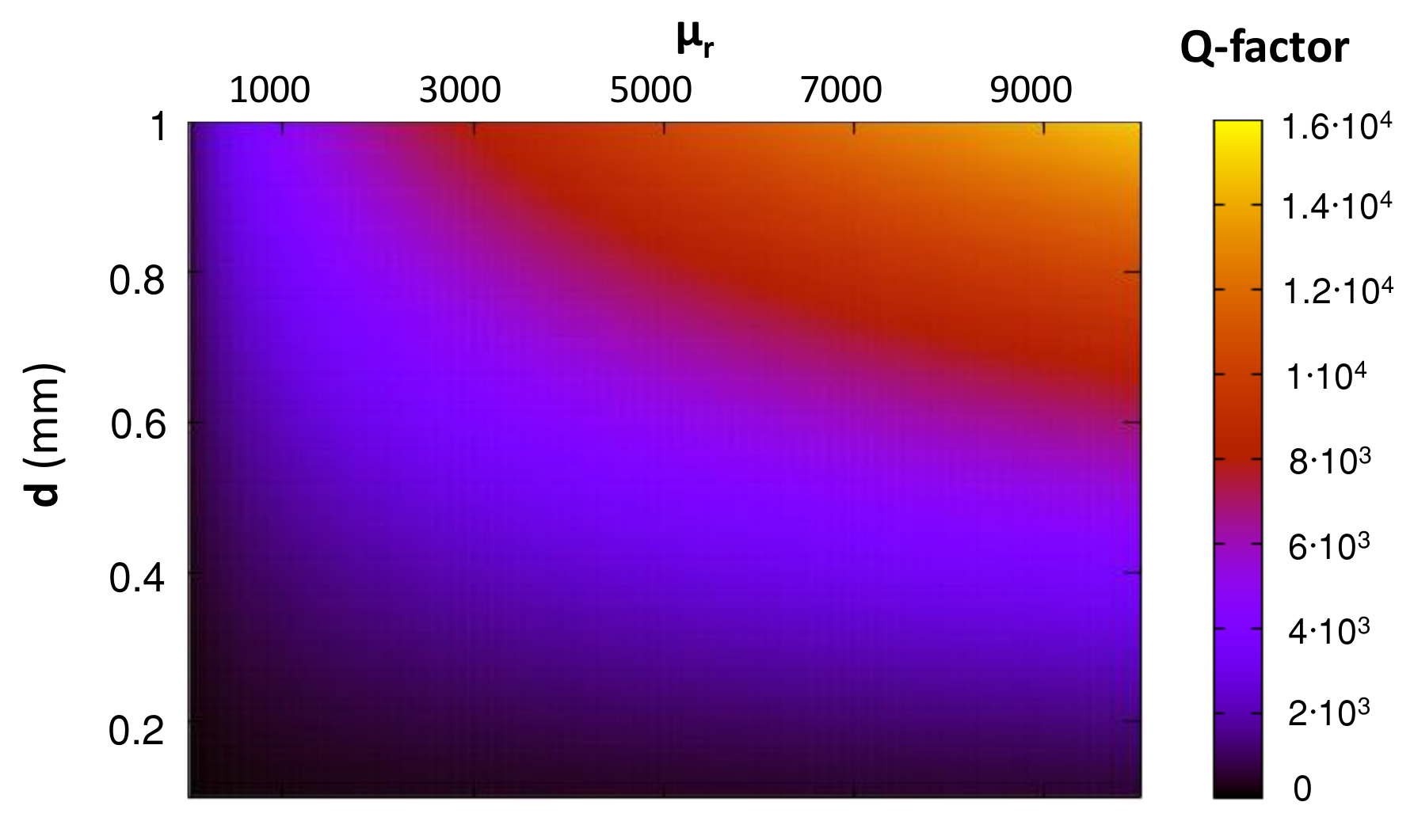}
  \caption{Map of Q-factor as function of $d$ and $\mu_r$. The other parameters are fixed at: $c=1$~uF, $\rho= 1.68\times 10^{-8}$, $N=10$. The resonant frequency over the entire map lies in the range $[1017.8,\,96610.3.7]$~Hz.
}
 \label{Fig:QvsDandMur}
\end{figure}

Practical constraints may require $\ell$ be smaller than  $1$~cm, in which case we need to consider thinner wires. Fig.~\ref{Fig:QvsCandD} reports a map of the Q-factor as function of $d$ and $c$. From this data, we can see that, even for wires as thin as $d= 0.5$~mm (hence $\ell=5$~mm), we can still obtain Q-factors of the order of $10^3$. Along the same lines, Fig.~\ref{Fig:QvsDandMur} reports a map of the Q factor as function of $\mu_r$ and $d$.

We end this section by proposing the following reasonable configuration, where the basic resonator consists of three capacitors $c_0= 3$~uF and a solenoid made out of a copper wire of diameter $d=1$~mm. The solenoid contains $N=10$ turns, has a diameter $D=5$~mm and is filled with an iron bar of $\mu_r=5000$. This configuration has a Q-factor of $6\times 10^3$ at a working frequency $\nu_0 =2.6$~kHz. Starting from this reference configuration, the valley-Chern effect can be generated by modifying the values of the capacitors to $c_\pm = (1 \pm r)\, c_0$, with $r = 0.13$,
 the optimal value found in section~\ref{Sec:DW}. This gives $c_+=3.4$~uF and $c_-=2.6$~uF.

\section{Conclusion}

In this work we introduced a novel platform based on inductively coupled discrete LC-resonators. In our proposal, the resonators are well defined and maintain their identities when incorporated in circuits. The couplings can be adjusted from weak to strong and from negative to positive, and their configurations can be arbitrarily complex. As such, the platform can be used to straightforwardly implement complicated tight-binding Hamiltonians, particularly, the topological Hamiltonians from the classification table \cite{ourpaper}.

The present work used this platform to implemente the valley-Chern effect, which supplied topological domain wall modes. In principle, the Berry curvature can be experimentally investigated by observing the dispersion of Gaussian wave packages \cite{GaoPRL2014,PricePRB2016}. However, the topology of the valley-Chern effect is best observed through its bulk-boundary correspondence. To some degree, the LC-circuit can be seen as a medium for electromagnetic wave propagation. Then, at the bulk-gap frequencies, the topological domain wall acts like a wave-guide for the electromagnetic waves. If, for example, a capacitor situated on the interface is excited with an ac-voltage of frequency inside the bulk-spectral gap, it will act like an antenna and the electromagnetic wave, instead of spreading throughout the whole LC-circuit, will propagate along the narrow wave-guide generated by the topological interface mode. As in the many previous applications mentioned in the introduction, the interface doesn't have to be straight but it can have, for example, a zigzagged shape or it can be reconfigured for various applications. 

There is a nice way to observe this phenomenon in a laboratory.  Specifically, suppose we insert an ac-driven light emitting diode (LED) in each of the elementary circuits, such that we can actually visualize the currents $i_{n,m}^\alpha$. If we pulse a capacitor located at the interface, and we modulate the pulse with a frequency inside the spectral gap, then we can visualize how this pulse propagates along the topological interface, with speeds that can be computed from the slopes of the chiral modes in Fig.~\ref{Fig:EdgeModes}. Furthermore, suppose that we excite the same capacitor, this time with an ac signal. As we sweep the frequency from zero and up, we should first observe LEDs being light up all across the LC-circuit, because the signal is spread by the bulk modes. But once the frequency enters the bulk spectral gap, we should see only the topological domain wall being light up. If we continue to increase the frequency, the signal will spread again throughout the LC-circuit.

Let us conclude by reminding that, before starting any experiments, one should map the landscape of the Berry curvature and the localization of the edge modes as function of parameters, in order to pin-point the optimal configuration of the LC-circuit. According to our estimates, even with reasonable configurations and materials, a Q-factor as high as $\times 10^4$ can be generated. Probing fundamental physics, such as the critical regime at a topological Anderson localization-delocalization transition, becomes feasible, and this makes the proposed platform very special.

\acknowledgments{ All authors acknowledge support from the W.M. Keck Foundation.}

\end{document}